\documentclass{vgtc}                          

\graphicspath{{figures/}{pictures/}{images/}{./}} 

\usepackage{times}                     

\usepackage{tabu}                      
\usepackage{booktabs}                  
\usepackage{lipsum}                    
\usepackage{mwe}                       

\usepackage{mathptmx}                  

\usepackage{graphicx}
\usepackage{array}
\usepackage{multirow}
\usepackage{fancyvrb}
\usepackage{gensymb}
\usepackage{colortbl}

\onlineid{1168}

\vgtccategory{Research}

\vgtcinsertpkg

\title{EmBARDiment: an Embodied AI Agent for Productivity in XR}

\author{Riccardo Bovo \\
\scriptsize  Imperial College London, UK
\and Steven Abreu \\
\scriptsize  University of Groningen, Netherlands
\and Karan Ahuja \\
\scriptsize Google, USA
\and Eric J Gonzalez \\
 \scriptsize Google, USA
\and  Li-Te Cheng \\
\scriptsize Google, USA
\and \\ Mar Gonzalez-Franco \\
\scriptsize Google, USA
}

\teaser{
  \centering
  \includegraphics[width=\linewidth]{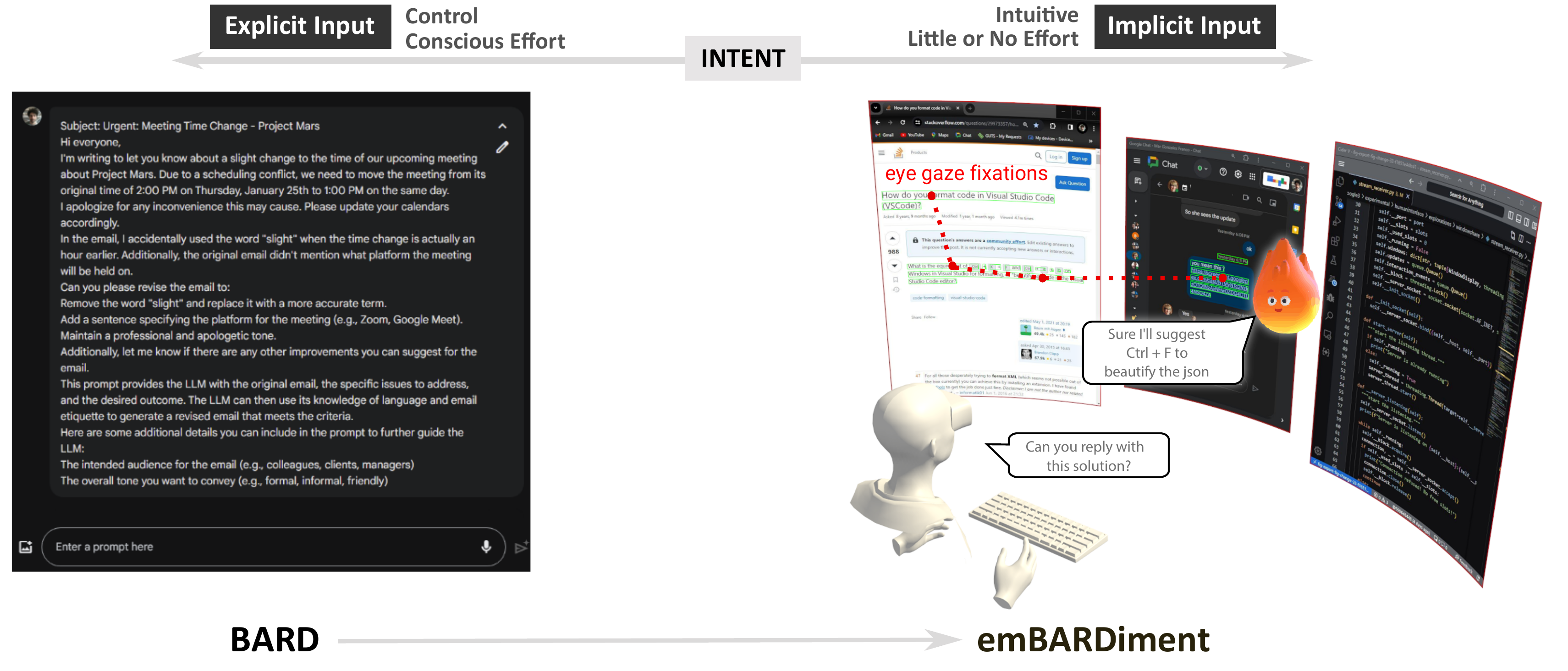}
  \caption{EmBARDiment introduces an attention framework that derives context implicitly from user eye-gaze, and contextual memory within the XR environment. It uses the implicitly information bundled with explicit verbal inputs to elicit grounded communication between the User and the AI Agent.}
  \label{fig:teaser}
}

\abstract{
   XR devices running chat-bots powered by Large Language Models (LLMs) have the to become always-on agents that enable much better productivity scenarios. Current screen based chat-bots do not take advantage of the the full-suite of natural inputs available in XR, including inward facing sensor data, instead they over-rely on explicit voice or text prompts, sometimes paired with multi-modal data dropped as part of the query. We propose a solution that leverages an attention framework that derives context implicitly from user actions, eye-gaze, and contextual memory within the XR environment. Our work  minimizes the need for engineered explicit prompts, fostering grounded and intuitive interactions that glean user insights for the chat-bot. }


\begin{document}


\firstsection{Introduction}

\maketitle

\begin{figure*}[t]
    \centering 
    \includegraphics[width=\textwidth]{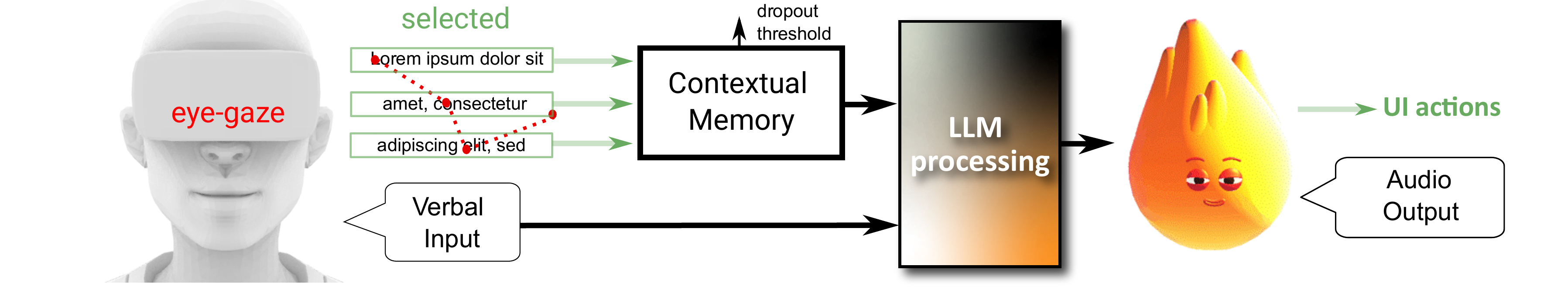}
    \caption{ Schema of EmBARDiment. The attention frameworks leverages implicit eye-gaze to select contextual information and bundles it with explicit verbal inputs. This elicits grounded communication between the User and the AI Agent.} 
    \label{fig:system} 
\end{figure*}

The rapid advancement of Large Language Models (LLMs) has revolutionized human-computer interaction, with chat-bots such as ChatGPT \cite{openai2022chatgpt}, Claude \cite{claudeai2023}, and BARD \cite{google2023bard} emerging as the primary interface for engaging with these powerful AI systems.  However, as LLMs continue to evolve, their potential extends beyond text-based interactions, particularly in the realm of extended reality (XR) environments. Consumer-grade XR headsets (e.g., Apple Vision Pro, Oculus Pro, Pico Neo Eye) uniquely offer eye tracking capabilities/support, which traditional screens lack at a comparable consumer level. Leveraging these XR-specific user facing sensors is crucial for future AI interfaces, as XR users increasingly interact with AI for productivity tasks or real-world attention applications. The integration of LLMs as embodied, always-on immersive agents within augmented and virtual reality headsets holds immense promise for transforming user experiences and enabling seamless, context-aware assistance.Despite the increasing sophistication of LLMs, current chat-bot implementations heavily rely on explicit voice or text prompts, often demanding multiple iterations to refine the desired tone and context for the optimal output, leading to lengthy interactions. In XR environments, this approach fails to capitalize on the diverse range of natural inputs available, as contemporary XR headsets (such as Meta Quest Pro \cite{metaquestpro2022} and Apple Vision Pro \cite{applevisionpro2022}) are equipped with inward-facing sensors capable of capturing rich user data. Although recent advancements in LLM architectures, such as Gemini 1.5 \cite{google2023gemini}, aim to support up to 10M tokens of context attached to a prompt, the current explicit input modalities in XR devices remain limited. For instance, speech input has an estimated universal throughput of only 39 bits per second \cite{coupe2019different}, while text input in XR technologies remains cumbersome and inefficient \cite{grubert2018text}. Our approach is particularly useful for XR environments due to their large field of view (FoV) displays, which can support multiple windows for productivity tasks \cite{biener2020breaking, pavanatto2021we}. While it is possible to provide all the data available on these windows to LLMs as context, this can result in a large amount of data, making the system less responsive, computationally heavy, and challenging to maintain nuanced conversations, as not all information may be relevant \cite{li2023improving,Jeuris2016}. Consequently, directly translating \textit{explicit} prompt-based chat-bots to embodied agents in XR is likely to be sub-optimal. To address these challenges, we propose EmBARDiment, a novel approach that leverages an \textit{implicit} attention framework in combination with a contextual memory to enable embodied LLM agents in XR environments (see Fig~\ref{fig:system}).  Our solution aims to implicitly derive context from the user's actions, eye-gaze, and visual saliency within the XR environment, minimizing the reliance on engineered explicit prompts. As a result, we facilitate highly targeted and intuitive interactions, allowing the agent to infer the user's intentions and needs based on their behavior and focus of attention, both present and past. For example, if a user is reading a document, the agent can access the content the user has read and utilize it as context for generating relevant responses, alongside the query prompt. Unlike prior works on gaze in XR that focus primarily on one-off visual querying  \cite{romaniak2020nimble,Miniotas2006SpeechGaze}, our approach combines gaze-saliency context with LLM-powered continued interaction, offering the best of both worlds. We hypothesize that this approach will enhance user interactions with the embodied agent, as it establishes a shared "theory of mind" and provides highly contextual assistance, potentially simplifying interactions to the level of natural language commands like "put that there" \cite{10.1145/800250.807503}.

\noindent The primary contributions of this work are as follows:

\begin{itemize}

\item We propose a novel attention framework leveraging gaze-based saliency driven contextual memory for embodied LLM agents in XR, thus enabling implicit user cues for context-aware assistance.

\item We conduct user studies to empirically evaluate the effectiveness of our proposed contextual framework in enhancing user interactions and experience, and showcase its efficacy over baseline explicit text-based input condition.

\end{itemize}

\section{Related work}

\subsection{Context-Aware Assistants for Productivity in XR}

The growing interest in XR multi-window environments for productivity \cite{biener2020breaking, pavanatto2021we} presents new opportunities and challenges for integrating context-aware assistants. These environments offer users cost-effective, customizable, and private workspaces, even on the go \cite{fereydooni2020virtual, ofek2020towards}. Moreover, XR multi-window environments can support remote work by mitigating issues like distractions, insufficient workspaces, and the struggle to separate work from personal life \cite{fereydooni2020virtual}, maybe even providing better multitasking solutions, as we can easily maintain multiple windows open at the same time \cite{pavanatto2021we}, even on mobile office work settings \cite{ofek2020towards}. The development of context-aware assistants \cite{dey2001conceptual} has seen a plethora of solutions, gleaning insights about the user (e.g. activity \cite{abowd1999towards}, pose \cite{schilit1995system}, preferences \cite{dey2000context}, task at hand \cite{dey1999conference}) and immediate environment (e.g. location \cite{mccarthy1999activemap}, ambient light  \cite{brumitt1999easyliving}). In the domain of XR, context-aware assistants have been explored to enhance education \cite{suzuki2023xr}, industrial training and maintenance \cite{greci2022xr} and enable more natural interactions \cite{GuoGRAVL-BERT:Resolution}. For instance, in the domain of immersive user experience, \cite{GuoGRAVL-BERT:Resolution} proposed a framework that utilises the metadata found in the field of view to help ground the conversation between human and AI agent \cite{GuoGRAVL-BERT:Resolution}.  While context-aware assistants in XR have been explored for various applications, there is a notable gap in research concerning their application for knowledge workers in XR environments, despite the recognized potential of XR for enhancing productivity work \cite{biener2020breaking,pavanatto2021we,fereydooni2020virtual}. Productivity within XR presents unique challenges that render traditional AI agent interfaces, such as chat-bots, inefficient, primarily due to the cumbersome and inefficient nature of typing in XR \cite{grubert2018text}. Additionally, alternative channels like speech, because of their intrinsic limitations \cite{coupe2019different}, are not deemed suitable to generate the complex prompt required for knowledge workers \cite{noy2023experimental}. Furthermore, the information accessible across multiple windows in XR environments, perhaps with even more screens open that in a regular PC, can pose difficulties for assistants in delivering targeted and efficient productivity support \cite{li2023improving}. EmBARDiment tackles these challenges by employing multimodal interactions and develops a gaze-driven contextual memory. This approach allows embodied AI agents to implicitly extract relevant context from the user's focus of attention within the XR environment, thereby facilitating more productive interactions.

\subsection{Gaze attention driven Multimodal XR Interactions}

Prior work in HCI leverages inputs such as gaze and pointing gestures concurrently with speech commands to generate multi-modal interactions \cite{pfeuffer2024design, gonzalez2024guidelines}. Richard A. Bolt's foundational study "Put-That-There" \cite{10.1145/800250.807503} examined the fusion of vocal instructions and manual gestures to improve engagement with graphical user interfaces. Likewise, \cite{Miniotas2006SpeechGaze} investigated the combination of eye tracking and verbal communication, particularly for engaging with small, densely arranged on-screen elements. Recent studies have shifted focus towards comprehending visual context and leveraging intuitive communication to improve interactions. Specifically, Sven Mayer et al.  \cite{Mayer2020EnhancingWorldGaze}  conducted a study that incorporated GPS location and user's head gaze into mobile voice assistants, enhancing their comprehension of nearby shops and buildings by understanding a user's location and visual direction, so that users can perform implicit verbal requests about their surroundings \cite{Mayer2020EnhancingWorldGaze}.  Karan Ahuja et al. \cite{ahuja2020direction} made use of speech as a direction communication channel akin to user-gaze to facilitate interactions across smart environment ecosystems.   Romaniak Yevhen et al. \cite{romaniak2020nimble} presented \textit{Nimble}, a mobile interface that combines visual question-answering frameworks with gesture recognition, focusing on directive gestures, to enhance user engagement. Sebastian Marwecki et al. \cite{marwecki2019mise} exploited the implicit attention cues of gaze in \textit{MiseUnseen} to facilitate novel interactions inside XR environments, including creating or arranging objects covertly at runtime. Riccardo Bovo et al. \cite{Bovo2023Speech-AugmentedAnalysis} exploited collaborative speech to improve accuracy head-based visual attention cue's. These works highlight the effectiveness of using visual attention or pointing gestures to aid in understanding the user's intended referent during verbal requests. However, these works primarily focus on providing directional information at the moment of the request. In contrast, EmBARDiment extends these techniques by implementing a contextual memory that encodes information about temporal gaze-based saliency, allowing the assistant to leverage this context up to the moment of the verbal request, thus enabling a more contextually relevant and efficient interaction paradigm.

\subsection{From Chat-bots to Embodied XR Agents}

Chat-bots have come a long way since their early days of rudimentary pattern matching, now incorporating sophisticated Natural Language Processing (NLP) techniques for nuanced text and voice interactions \cite{adamopoulou2020chatbots}. Advancements in NLP have enabled these systems to comprehend complex inputs and generate contextually relevant responses \cite{Chakraborty2023}. This evolution has transformed user interactions with AI, shifting from typing to voice-based interfaces \cite{malik2021automatic, radford2023robust, tan2021survey}, reflecting the importance of creating communication that mirrors human-human interactions \cite{poushneh2021humanizing, chitto2020automatic}. The recent debut of powerful LLMs like ChatGPT, BARD, and Gemini has sparked renewed interest in AI-driven chat-bot applications \cite{Chakraborty2023}, leading to increased user engagement and scientific research \cite{chakraborty2023overview}. This renewed interest has also motivated the integration of AI chat-bots and agents into XR applications, in the form of embodied agents \cite{hartholt2019ubiquitous, numan2023ubiq}. These embodied agents aim to further enhance user engagement and natural interaction through visual and behavioral cues. EmBARDiment builds upon this renewed interest in embodied XR agents, and introduces an attention based framework that dynamically adjusts to the user's current activity and context within the XR environment. 

\section{EmBARDiment}

EmBARDiment is a practical example of our framework that is implemented as an XR application that seamlessly integrates speech-to-text, text-to-speech, gaze-driven saliency and LLMs to enable a voice-interactive AI agent within a multi-window XR environment. In this section, we describe the system architecture (Fig~\ref{fig:system}), its key components and their integration. The application was developed using Unity \cite{Unity2023} and deployed on the Oculus Quest Pro \cite{metaquestpro2022}. 
The code is available at the following GitHub repository: \href{https://emBARDiment.github.io}{https://emBARDiment.github.io}.

\subsection{Embodied AI Agent}

EmBARDiment features an embodied AI agent that serves as the primary anchor and interface for user interaction. The AI agent is embodied into a cute avatar designed to provide an engaging and intuitive experience by combining verbal and non-verbal cues. When the AI agent receives a response from the ChatGPT-4 API, it uses the Google Cloud Text-to-Speech API to generate speech and corresponding visemes for lip synchronization and facial animations. This integration creates a more lifelike and immersive experience for the user, mimicking human-like interactions. The AI agent's embodiment plays a crucial role in establishing a sense of presence and facilitating natural communication. By leveraging the user's gaze and saliency history, the AI agent can move around the different windows showcasing visually a level of shared understanding of the user's current focus.

\subsection{Multimodal Interaction}

EmBARDiment leverages multiple input and output modalities to enable seamless communication between the user and the embodied AI agent \cite{pfeuffer2024design, gonzalez2024guidelines}. The user can initiate a verbal request by pressing a key and then speaking ('V' on the keyboard). The user's speech is converted to text using the Google Speech-to-Text API \cite{GoogleCloudSpeechToText}, which sends the audio data to the Google cloud and returns the transcribed text that is then displayed on the fly in a UI panel beneath the AI agent, providing visual feedback to the user (see supplementary video).
Once the final transcription is received from the Google Speech-to-Text API, it is processed by any LLM API for natural language understanding and generation. The system maintains a chat history to preserve the interaction context, allowing the AI agent to generate relevant and coherent responses. The response is then displayed on a UI panel beneath the AI agent, and we use again the Google Cloud Text-to-Speech API \cite{GoogleCloudTextToSpeech} but now to convert the text into speech. This process also produces phonemes and their corresponding visemes, which are used to animate the AI agent's facial expressions and lip movements in sync with the spoken words \cite{volonte2022headbox}.

\subsection{Gaze-Driven Contextual Memory}
EmBARDiment builds on top and extends an existing open-source multi-window XR environment WindowMirror that captures existing windows from a PC and renders them inside the XR environment \cite{bovo2024windowmirror}. EmBARDiment processes each window frame using the Google Vision API \cite{GoogleVisionOCRAPI} to perform optical character recognition (OCR), extracting the text content and its position within the frame. By correlating the spatial position of the text with the eye-tracking data from the XR headset, EmBARDiment can determine which text the user is currently paying attention to. The system maintains a buffer of the user's saliency history, preserving the order of the text to ensure coherence. The system only considers fixations, which are determined by a spatial and temporal threshold adapted specifically to the user’s reading task. We consider a fixation if the gaze spends over 120 millisecond on top of a single word’s bounding box. The temporal threshold of 120ms is considered the minimum time required for effective visual information processing during reading \cite{rayner1998eye}. The contextual memory in our system has a maximum capacity of 250 words to enable basic episodic memory, rather than long term memory. This capacity was defined during the pilot studies. For context, our texts ranged from 240 to 400 words in length, and participants took approximately one minute to read each text in full.. Our agent contextual memory works by simply discarding older information as the user focuses on new content. When the user makes a verbal request, the contextual memory is combined with the user's query and sent to the ChatGPT-4 completion API for processing. This approach allows the AI agent to generate responses that are grounded in the user's current focus and saliency history. After each request, the contextual memory buffer is cleared. Below you can see an example of the generated Prompt containing the user verbal request , some prompt engineering and the gaze-driven contextual memory:

\definecolor{darkgreen}{rgb}{0.0, 0.39, 0.0} 
\definecolor{darkyellow}{rgb}{0.7,0.7,0.3} 
\definecolor{darkcyan}{rgb}{0.0,0.5,0.5} 

\begin{Verbatim}[commandchars=\\\{\}]
\textcolor{darkyellow}{User's verbal request}
\textcolor{darkgreen}{Prompt Engineering} 
\textcolor{darkcyan}{Contextual information}

\{
    \small "role": "user",
    \small "message": \textcolor{darkyellow}{user verbal request from speech-to-text.} 
    \small \textcolor{darkgreen}{Below is a dataset representing my visual attention it} 
    \small \textcolor{darkgreen}{contains the text i have been reading from the windows} 
    \small \textcolor{darkgreen}{I have been observing. Please use this information to} 
    \small \textcolor{darkgreen}{inform your response to my request.} 
    \small \textcolor{darkcyan}{window_id_1: \{gaze-selected text on window_id_1\},} 
    \small \textcolor{darkcyan}{window_id_2: \{gaze-selected text on window_id_2\}.}
    \small \textcolor{darkgreen}{Respond with 6 sentences max, keep it under 400}
    \small \textcolor{darkgreen}{characters maximum.".}
\}
\end{Verbatim}

\begin{figure*}
   \centering 
    \includegraphics[width=1\textwidth]{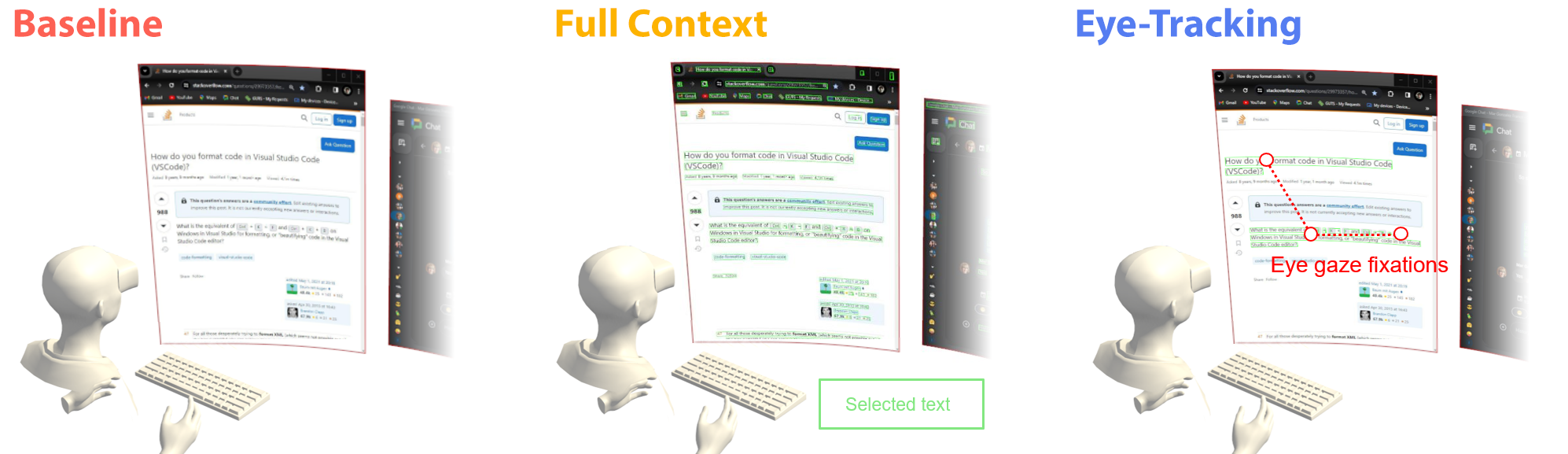} 
    \caption{Experiment Conditions.  (A) Baseline:  no contextual information selected. (B) Full-Context: All the contextual information are selected. (C) Eye-gaze: information are selected based on eye-gaze fixations.} 
    \label{fig:conditions} 
\end{figure*}



\section{Experiment}
To evaluate EmBARDiment, we conducted a user study. Our evaluation aims to determine (i) the merit of visual attention as a method for selecting contextual information, and (ii) to evaluate the different options for agent embodiment and the positioning of the agent. 

\subsection{Participants}

We recruited 15 participants (10 male, 6 female, ${M}_{{Age}}$=31.4, ${SD}_{{Age}}$=5.9  years). Since in the experiment we use speech-to-text technologies we evaluated English speaking proficiency: on a scale from 1 (poor) to 5 (excellent), resulting in an average rating of ${M}_{{English}}$=4.81  ${SD}_{{English}}$=0.425. note that for 10 out of 15 participants English was their primary language. Additionally, since in the experiment we use eye-tracking technologies we also evaluated vision. In terms of vision correction, 5 participants reported having normal vision, while all the others had vision corrected to normal; 6 wearing glasses and 4 contacts.


\subsection{Design}

We designed a within-participant experiment with 3 conditions and 3 tasks, each participant completed 3 tasks. A Latin square design balanced the assignment of the 9 unique task-condition combinations, ensuring equal representation across 15 participants. Although presentation order was not fully counterbalanced, each condition appeared in each ordinal position (first, second, third) an equal number of times.
The 3 conditions:
 
\begin{itemize}
 \item A \textcolor{red}{Baseline} condition in which there is no contextual information added (see Fig~\ref{fig:conditions}a).
 \item A \textcolor{yellow}{Full Context} condition where all the contextual information is selected independently of the user's visual attention (see Fig~\ref{fig:conditions}b).
 \item An \textcolor{blue}{ Eye-Tracking} condition which selects information based on eye-gaze data (see Fig~\ref{fig:conditions}c).
\end{itemize}
 
The quantitative and qualitative comparison between these conditions will establish if and how contextual episodic memory will enable more implicit communication and better understand the role that eye gaze will play with AI agents in XR.

\begin{figure}[h!]
    \centering 
    \includegraphics[width=0.47 \textwidth]{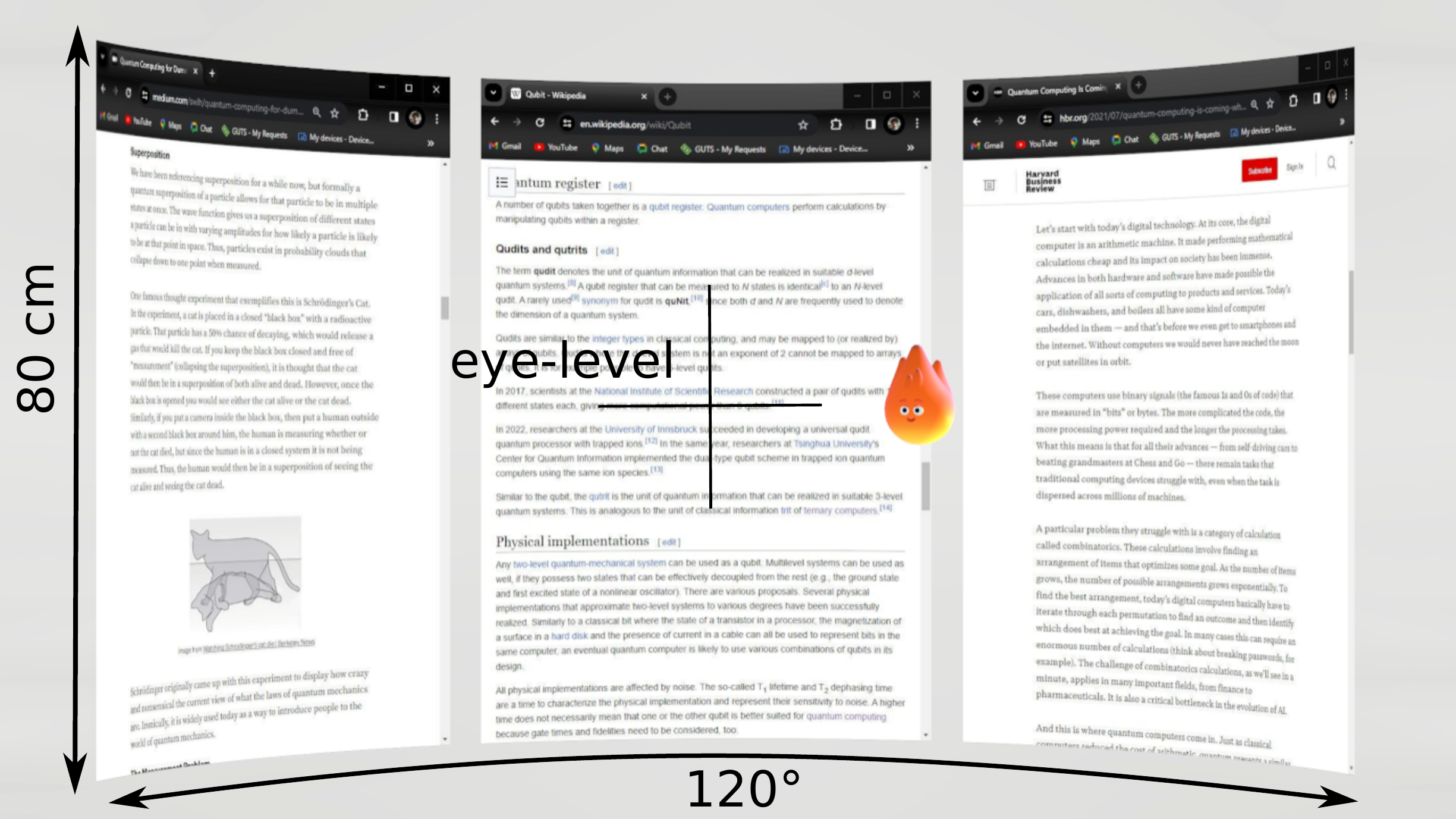} 
    \caption{Experiment layout, as seen by the participants during the experiment. The 3 texts are fixed for all participants to maintain stimuli consistency. The layout of windows spawned 1 meter away from the user's head, spanning 120\degree  (60\degree on the left and 60\degree on the right), and a resolution of 700x1200 px. A 3D model of the layout with reference textures is available at the following GitHub repository: \href{https://emBARDiment.github.io}{https://emBARDiment.github.io}.}
    \label{fig:layout} 
\end{figure}

\subsubsection{Q\&A Reading Task}
\label{sec:QA}
The three bodies of text of 270 to 400 words each, were selected for the experiment. All texts had a common thematic of Quantum Computing. The first was based on a Medium article about the "Schrödinger's Cat" \footnote{https://medium.com/swlh/quantum-computing-for-dummies-part-1-2686b9ba3c51}; the second was a Harvard Business Review article: "Quantum Computing Is Coming"\footnote{ https://hbr.org/2021/07/quantum-computing-is-coming-what-can-it-do}, and the third was a Wikipedia article discussing "Qubits" \footnote{{https://en.wikipedia.org/wiki/Qubit}}. The two questions regarding each reading were intentionally designed to be implicitly related to the context of the text and essentially meant to ask clarifications on what the participant had just read.


\begin{itemize}
    \item [cat] Questions for the "Schrödinger's Cat" text:
    \begin{enumerate}
        \item Why is it set to 50\%?
        \item I am confused, is the cat alive or dead?
    \end{enumerate}
    \item [HBR] Questions for the "Quantum Computing Is Coming" text:
    \begin{enumerate}
        \item What are the capabilities of computers nowadays?
        \item What problems do traditional computing devices struggle with?
    \end{enumerate}
    \item [wiki] Questions for the "Qudits and qutrits" text:
    \begin{enumerate}
        \item What letter can I use to denote the dimension of the system?
        \item I am confused about how many levels are possible?
    \end{enumerate}
\end{itemize}


\subsubsection{Multi Window Layout}
The window layout, as shown in Fig~\ref{fig:layout}, is an essential part of the task. The layout is intended to mimic aspects of productivity work, such as multitasking, by having multiple application windows open simultaneously. This setup mirrors common scenarios in physical multi-screen environments where multiple windows are open and used at the same time by users. 

\subsection{Procedure}
Participants received a project overview, consent form, demographic questionnaire, and experiment instructions. The experiment began with a reminder of their right to withdraw and the opportunity to seek clarifications. Our institution does not run an Ethics committee but does intensive training for employees and interns on to how to perform user studies ethically. All the authors running studies did complete the training.

\subsubsection{(Part 1) Q\&A Paper Reading:}
\label{sec:p1}
We asked participants to read the selected texts on paper. We then presented two questions about each text . This procedure was repeated for all three bodies of text (Section~\ref{sec:QA}). This initial step was used to make participants familiar with the questions and relevant information within the texts.

\subsubsection{(Part 2) XR Onboarding:}

After answering the questions, participants engaged in a similar task but this time inside XR and interrogating the AI Agent. To guarantee a successful experiment, participants were helped with the XR headset and were guided to perform eye-tracking calibration and testing. Participants were then asked to open the XR application and familiarize themselves with the modality of interaction with the AI Agent (i.e., button press and verbal request).

\subsubsection{(Part 3) Q\&A AI Agent Text Interrogation:}
\label{sec:p3}
Then, in the XR experience, they were again presented with the same 3 texts, each one visible on one of the 3 screens as shown in Figure~\label{fig:layout}. They were asked to read the text till they reach the text part pertinent to the question rehearsed on Part 1. Then they had to ask the questions to the AI agent and keep asking until the agent produced a satisfactory response, i.e. a response equivalent to their own early reply outside XR. We instructed participants to maintain the original implicit formulation of their questions in the first attempt inside XR. If they did not receive a relevant answer, they were allowed to reformulate their question up to five times in an attempt to generate a contextually meaningful response. After five attempts, the experimenter would prompt them to move on. The study was structured around the different experimental conditions (baseline, full context, eye-tracking). And each condition had one of the three reading tasks (i.e. one of the screens) randomly assigned (cat, HBR, wiki), so they varied by participants.

\subsubsection{(Part 4) Experience Survey:}
Following each completed condition, participants were asked questions from a subset of the Human Language Model Interaction Questionnaire (HLMIQ) \cite{lee2022evaluating} shown in Table \ref{tab:evaluation}, and gave verbal responses while the experimenter annotated those responses. 

\subsubsection{(Part 5) Agent Embodiment Preferences:}
Upon completion of the Q\&A tasks/survey, the study transitioned to the exploration of AI agent embodiment preferences. Participants were introduced to different possible embodiment types (no embodiment, non-anthropomorphic, or anthropomorphic embodiment), as shown in Figure~\ref{fig:embodiment_type}, and embodiment location (general fixed, or contextual positioning such as following the active window), as shown in Figure~\ref{fig:embodiment_position}. Participants were asked to engage with each option to form a basis for their preferred choice ( Table \ref{tab:preferences}), and they could change the options using the joystick on the XR controller.  The experiment structure allow participants to first acquire a grounding on the experience of interaction with the agent before expressing preferences about its embodiment, and position in the multi-window environment.

\subsection{Questionnaires}

To better characterize the user experience beyond the quantitative analysis of the attempts, we asked participants to fill a subset (Table ~\ref{tab:evaluation}) of the HLMIq  \cite{lee2022evaluating}. The HLMIq assesses important aspects such as: Helpfulness, Ease, Enjoyment, Satisfaction, Responsiveness of the AI Agent.After each experimental block consisting of the 3 trials/texts, each participant was shown a few options in terms of Embodiment Type and Location within the multi-window space and then completed a survey on (Table~\ref{tab:preferences}). 

\begin{table}[h!]
\centering
\caption{Human Language Model Interaction Questionnaire}
\label{tab:evaluation}
\begin{tabular}{|m{0.18\linewidth}|m{0.555\linewidth}|m{0.12\linewidth}|}
\hline
\textbf{Aspect} & \textbf{Question} & \textbf{Answ. Type} \\
\hline
\multirow{2}{*}{Helpfulness} & Independent of its fluency, how helpful was your AI Teammate in answering the questions? & Likert \\
\cline{2-3}
& Why did you find the AI Teammate helpful or unhelpful? Give a concrete example if possible. & Open \\
\hline
Ease & How easy was it to interact with the AI Teammate? & Likert \\
\hline
Enjoyment & How enjoyable was it to interact with the AI Teammate? & Likert \\
\hline
Satisfaction & I am satisfied with the answer I received? & Likert \\
\hline
Change & Did the way you chose to interact with the AI Teammate change after the first question? If so, how? & Open \\
\hline
Description & What adjectives would you use to describe the AI Teammate? & Open \\
\hline
Responsiv. & How responsive was the system? & Likert \\
\hline
\end{tabular}
\end{table}

\begin{table}[h!]
\centering
\caption{Survey on Agent Embodiment Preferences}
\label{tab:preferences}
\begin{tabular}{|m{0.18\linewidth}|m{0.33\linewidth}|m{0.35\linewidth}|}
\hline
\textbf{Aspect} & \textbf{Question} & \textbf{Possible Answers} \\
\hline
\multirow{2}{*}{Embodiment} & What of these 3 options do you prefer? & 1)Lack of embodiment 2) Non-anthropomorph. 3)Anthropomorph.\\
\cline{2-3}
   & Why did you choose the embodiment type? &  (Open ended)\\
\hline
\multirow{2}{*}{Location} & What of these 2 options do you prefer? & (1) Context-following, (2) Fixed  \\
\cline{2-3}
& Why did you choose the location? & (Open ended) \\
\hline
\end{tabular}
\end{table}

\section{Results}
Each of the 15 participants had to ask the AI agent 2 questions across the 3 conditions leading to a total of 90 questions. The potential attempts to reach satisfactory answers can range from red 90 to 450 (as each participant could reformulate the same question 4 times). In total we collected 184 attempts on the experimental setting. After each of the 3 conditions, participants completed Human Language Model Interaction questionnaires leading to 45 questionnaires. Participants also completed the agent embodiment questionnaires, leading to 15 filled surveys.

\subsection{Question Attempts}

\begin{figure}[h!]
    \centering 
    \includegraphics[width=0.9\linewidth]{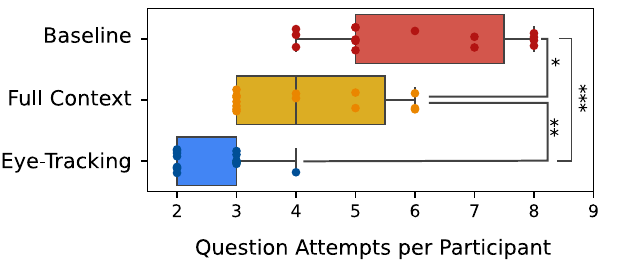} 
    \vspace{-4mm}
    \caption{ Box plot comparing participants' attempts. Each participant answered two questions, with up to five attempts per question. Significant differences highlighted: {$ $ * p $<$ .05, ** p $<$ .01, *** p $<$ .001}. } 
    \label{fig:attempts_number} 
\end{figure}

In our experiment, participants interrogate an AI agent, in search for explanations related to a featured text. If the AI's response is irrelevant or unacceptable, participants reformulate their question, a maximum of five times to obtain an acceptable answer. Fewer attempts indicate better performance, as they signify less effort to reach an acceptable answer. We analyzed the dialogues to count the number of attempts, totaling 184 verbal requests. To compare the number of attempts across conditions, we performed a repeated measures ANOVA with Condition as a three levels factor (baseline, full context, eye-tracking) . Fig~\ref{fig:attempts_number}.

\begin{figure}[h!]
    \centering 
    \includegraphics[width=0.5\textwidth]{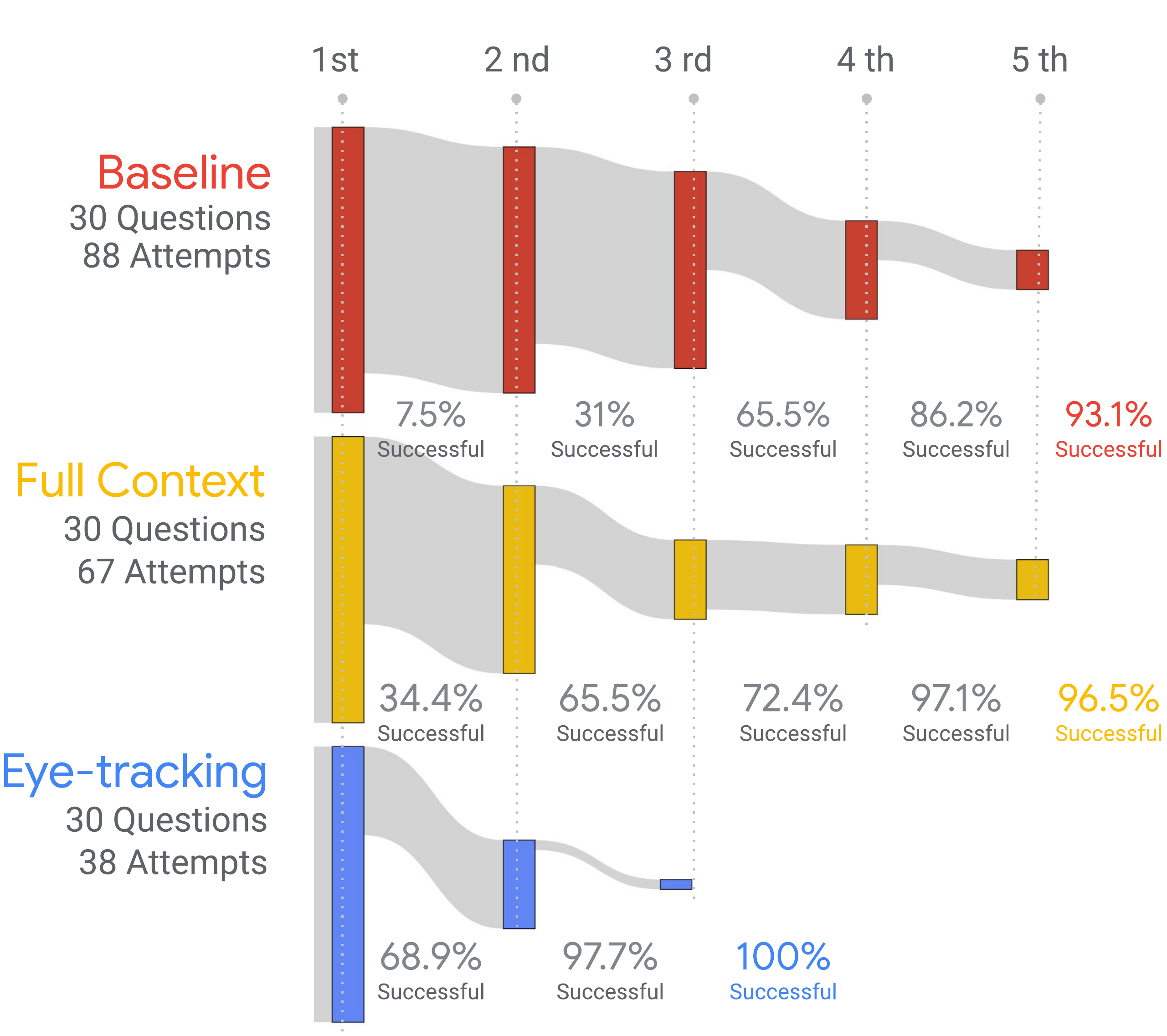} 
    \caption{Sankey diagram depicting the success rate at each subsequent attempt (i.e., up to five attempts per question). } 
    \label{fig:attempts_success_rate} 
\end{figure}

\noindent The ANOVA on the number of questions attempts indicate a significant difference between the conditions ($F(1.428, 19.993) = 26.416$,  $p< .001$), Greenhouse-Geisser correction was applied. To further unpack the results, we conducted pairwise post-hoc comparisons using dependent-samples t-tests with a Holm correction, revealing significant differences across all conditions. The comparison between the number of attempts in the Baseline ($M = 5.93$, $SD = 1.58$) and  Full Context ($M = 4.20$, $SD = 1.32$) revealed a significant difference of ($MD=1.73, p_{Holm} = 0.012$) and a large effect size (Cohen's d = 1.393), indicating a significant reduction in the number of attempts in the Full Context condition. This reduction in attempts is further emphasized in the comparison between the Full Context and  Eye-Tracking ($M = 2.53$, $SD = 0.64$), which showed a significant difference of ($MD=1.67, p_{Holm} = 0.001$) and an large  effect size (Cohen's d = 1.339), demonstrating a further decrease in number of attempts for the Eye-Tracking condition. To further clarify the results, we utilized a Sankey plot to illustrate the success rates at each attempt (Fig~\ref{fig:attempts_success_rate}). In the Eye-Tracking condition,over 90\% of participants achieved the intended result on the second attempt. In the Full Context condition, over 90\% succeeded by the fourth attempt. In contrast, in the Baseline condition, 90\% of participants reached a satisfactory result only by the fifth attempt.

\subsection{HLMIq}
\label{sec:questionnaire}
Participants completed HLMIq questionnaire after each experimental condition, resulting in a total of 9 entries per condition and 45 overall. We used the Aligned Rank Transform (ART) ANOVA \cite{wobbrock2011aligned}, which accommodates within-subject designs and ordinal data, such as Likert-scale responses. If a significant effect was detected, we conducted pairwise post-hoc comparisons using Tukey's Honestly Significant Difference (HSD) test on ART-transformed data (ART-lm).\footnote{We used the ARTool implementation for R: \url{https://github.com/mjskay/ARTool}, based on the original method: \url{https://depts.washington.edu/acelab/proj/art/}.} Significant effects were found for Helpfulness and Satisfaction, while no significant results emerged for Ease, Enjoyment and Responsiveness. For conciseness, we only report the analysis of the former.

\subsubsection{Satisfaction (Likert)}
The ART revealed a significant on Satisfaction scores ($F(2, 28) = 15.565$, $p < 0.001$). Post-hoc pairwise comparisons between Eye-Tracking (${Mdn} = 5$, ${IQR} = 1.0$) and Baseline (${Mdn} = 3$, ${IQR} = 2.0$) highlight that Eye-Tracking had significantly higher Satisfaction scores ($p < 0.0001$). Similarly, post-hoc pairwise comparisons between Eye-Tracking and Full Context (${Mdn} = 4$, ${IQR} = 0.5$) highlight that Eye-Tracking had significantly higher Satisfaction scores ($p = 0.0102$). Although Baseline and Full Context differed in median scores, the difference did not reach statistical significance ($p = 0.0585$).

\begin{figure}[h!]
    \centering 
    \includegraphics[width=0.46\textwidth]{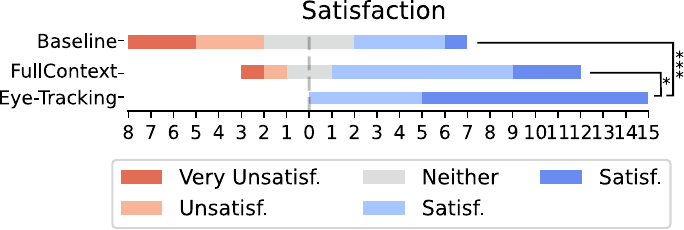} 
    \caption{Satisfaction responses to the HLMIq. Significant differences highlighted with: {$ $ * p $<$ .05, ** p $<$ .01, *** p $<$ .001}.} 
    \label{fig:satisfaction} 
\end{figure}

\subsubsection{Helpfulness (Likert)}
The ART revealed a significant effect on Helpfulness scores ($F(2, 28) = 14.787$, $p < 0.001$). Post-hoc pairwise comparisons between Eye-Tracking (${Mdn} = 5$, ${IQR} = 0.5$) and Baseline (${Mdn} = 3$, ${IQR} = 2.0$) show that Eye-Tracking was rated significantly higher ($p < 0.0001$). Similarly, comparisons between Eye-Tracking and Full Context (${Mdn} = 4$, ${IQR} = 1.0$) reveal that Eye-Tracking demonstrated significantly higher helpfulness ($p = 0.0079$). However, no significant difference was found between Baseline and Full Context ($p = 0.1007$). These results are consistent with the findings from our quantitative analysis of question attempts.

\begin{figure}[h!]
    \centering 
    \includegraphics[width=0.46\textwidth]{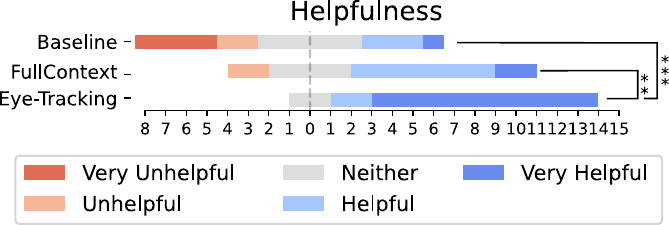} 
    \caption{Helpfulness responses to the HLMIq. Significant differences highlighted with: {$ $ * p $<$ .05, ** p $<$ .01, *** p $<$ .001}.} 
    \label{fig:helpfulness} 
\end{figure}

\subsubsection{Helpfulness (Open)}
To further unpack this result we tap on the Open ended question of the HLMIq:``Why did you find the AI Teammate helpful or unhelpful?''  For the Baseline condition, participants noted that the initial answers of the AI Agent were general and unrelated to the context of the text and that a satisfactory answer was only obtained after significant prompting.  \textit{P5: “..it did not give any specific information to the text i was reading..”}   \textit{P8: “..took a lot of prompting for the first answer but did really good in the second question..”} For the Full Context condition, participants experienced mixed results saying that the journey to the correct answer was often cumbersome \textit{P3: “..it was able to answer my question even if it took multiple attempts.”} \textit{P5: “..it's giving a very generic answer, unless I asked twice and be specific.”} In the eye-tracking condition, participants reported accurate answers straight to the point. \textit{P4: “..it answered the question very accurately.”} \textit{P9: “..it answered the questions straight away.”}

\subsubsection{Behavior changes (Open)}
\label{sec:changes}
To better understand the differences between conditions we also looked at the quesiton: ``Did the way you chose to interact with the AI Teammate changed after the first question? If so, how?''.  In the Baseline condition the majority of responses indicate participants provided more information in an attempt to guide the AI. \textit{P6: “I gave explicit content from the article to ensure it gave me the right answer.”}, \textit{P9: “I gave more context to it.”} Participants reported learning from their initial interaction, indicating an iterative process of adjusting their approach based on the AI's responses.
In the Full Context condition the majority of participant still reported the same strategies “of adding more information”. 

\subsection{Conversation Evaluation}
We review the recorded conversation to better characterize the differences between the Full Context and Baseline condition. And to understand how and why the Full Context condition failed in answering correctly. 

\subsubsection{Conversation Examples}
\label{sec:convexamples}
To further support this section we report two full example conversations from participant P2. Similarly to the HLMIq we also find two strategies in these examples:

\definecolor{darkgreen}{rgb}{0.0, 0.39, 0.0} 
\definecolor{darkred}{rgb}{0.55, 0.0, 0.0} 

\begin{Verbatim}[commandchars=\\\{\}]
\colorbox{yellow}{\small Strategy 1: Referring to Context} 
\colorbox{green}{\small Strategy 2: Adding More Information}

\textbf{\small Example 1 (P2) Baseline:}

\small Participant 2 : \small  Why is the chance of radioactive
\small                 decay 50\%?
\small AI Agent : \textcolor{darkred}{\small The chance of radioactive decay is not}
\small            \textcolor{darkred}{\small inherently 50\% for all isotopes...}
\small Participant 2 :\colorbox{yellow}{\small In this example}, though. Why is it 50\%?
\small AI Agent : \textcolor{darkred}{\small In your example, if the chance of decay} 
\small            \textcolor{darkred}{\small is 50\%, it means we're looking at a }
\small            \textcolor{darkred}{\small specific moment..}
\small Participant 2 :\colorbox{green}{\small In the Schrödinger's cat example}, 
                \small why is the chance of decaying 50\%?
\small AI Agent : \textcolor{darkgreen}{\small In Schrödinger's cat thought experiment,}
\small            \textcolor{darkgreen}{\small the 50\% decay chance is a simplification}
\small            \textcolor{darkgreen}{\small to illustrate quantum superposition..}

\textbf{\small Example 2 (P2) Full Context:}

\small Participant 2 : What are the core capabilities of 
\small                 nowadays computers?
\small AI Agent : \textcolor{darkred}{\small Quantum computers harness the}
\small            \textcolor{darkred}{\small principles of quantum mechanics to}
\small            \textcolor{darkred}{\small process information....}
\small Participant 2 : What are the core capability of 
\small                 computers\colorbox{yellow}{\small described in this article}?
\small AI Agent : \textcolor{darkgreen}{\small The article highlights that digital}
\small            \textcolor{darkgreen}{\small computers excel at arithmetic}
\small            \textcolor{darkgreen}{\small operations…}

\textbf{\small Example 3 (P2) Eye-Tracking:}

\small User : What's the term that describes the capabilities of 
\small        computers?
\small AI Agent : \textcolor{darkgreen}{\small The term that describes the capabilities}
\small            \textcolor{darkgreen}{\small of computers is arithmetical machines.}
               
\end{Verbatim}

\subsection{Agent Embodiment Preferences}
This Agent Embodiment Preferences Survey was once by each participant at the end of the whole procedure (n=9).
\subsubsection{Agent Embodiment Type}
\begin{figure}[h!]
    \centering 
    \includegraphics[width=0.8\linewidth]{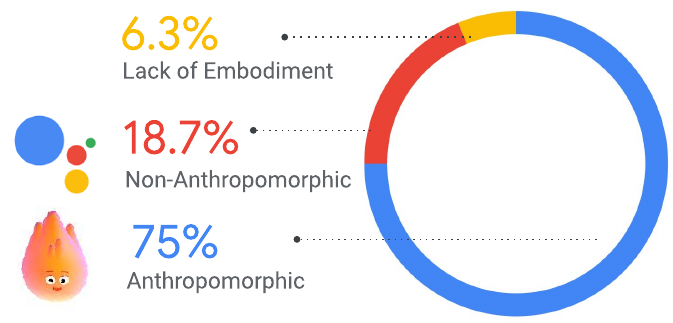} 
    \caption{Circle chart representing user's preferences in terms of embodiment's type.} 
    \label{fig:embodiment_type} 
\end{figure}

When we asked, ``How do you prefer the visual representation of a digital assistant to be designed?'' Participants were shown three options: Lack of Embodiment, Anthropomorphic, and Non-anthropomorphic.  After experiencing each of the conditions, participants were asked to express a preference. The option that received the most preferences was Anthropomorphic (75\%), followed by Non-Anthropomorphic (18.7\%) and Lack of Embodiment (6.3\%) (Figure ~\ref{fig:embodiment_type}).

To further unpack these results, participants were asked to justify their preferences. Those whose preference was anthropomorphic mentioned a desire for interactions that mimic human communication.  \textit{P6: “A human-like agent makes me a bit less self-conscious about making a verbal request”}. However, many who expressed a preference for the anthropomorphic design were flexible about the representation; they would be fine with an abstract, non-anthropomorphic option, as long as it was visually present. \textit{P1: “I don't mind between the two  as long as it is represented. I don't want the disembodied because I don't know if it's present or listening to me.”} The representation in a visual form was signifying to participants the assistant's presence and attentiveness. \textit{P5: “I like it to be there because it reminds me that I can ask questions”} Those who chose lack of embodiment valued the absence of representation, as it was non-distracting and not visually imposing. 

\subsubsection{Agent Embodiment Location}

\begin{figure}[h!]
    \centering 
    \includegraphics[width=\linewidth]{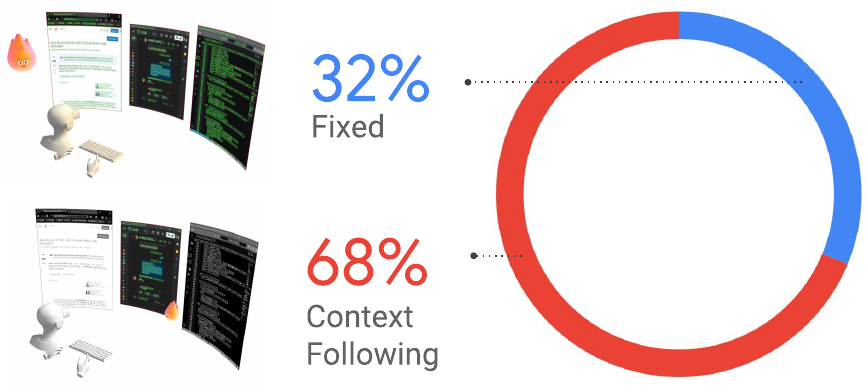} 
    \caption{Circle chart representing user's preferences in terms of embodiment's location. } 
    \label{fig:embodiment_position} 
\end{figure}

In a similar way we then asked participants, ``Where do you prefer the AI assistant to be located?'' giving them two choices: Context-following or Fixed. After showing these options to participants and addressing any questions, they were asked to make a choice. Most participants preferred Context-following (68\%), but around one-third opted for Fixed (32\%) (Fig~\ref{fig:embodiment_position}).

When asked To justify Context-following choices participant mentioned a desire for the digital assistant to be easily accessible without requiring the user to turn their head or attention.\textit{P2: “I don't have to turn my head. ”} \textit{P7: “I would want the avatar to follow my gaze better. ”} The context-following group also highlighted a desire for clarity in how the assistant's movements and positioning reflect its awareness of the context. \textit{P1: “Context following makes it easier for me to understand what the avatar is aware of.”}

\section{Discussion}

\subsection{Why the AI Agent failed in Full Context Condition}
We further look at the results and we find a general consistent trend: the AI Agent failing on the Full Context condition. But what are the differences between the Full Context and the Baseline conditions that cause this? One of our main learning was that the AI Agent performance faltered in the Full Context condition when compared to the Eye-Tracking condition, we believe this is due to the expansive context, which allowed for multiple potential answers. This was in stark comparison to our 250 word episodic memory implemented for gaze. For instance, if we look at the Conversation examples (section  \ref{sec:convexamples}), in the "Example 2 (P2) Full Context", to the question \textit{"What are the core capabilities of nowadays computers?"} the AI Agent answer \textit{"Quantum computers harness the principles of quantum mechanics to process information"} is influenced by the broader context about quantum computing, as all 3 articles talk about quantum computing while only one talks about the capabilities of computers nowadays. Given that the Full Context is primarily concerned about quantum computing, the AI Agent tended to focus its answers on quantum computers, illustrating how extensive context can skew its responses. This led to inaccuracies, as the model struggled to integrate the question on the specific context the user was reading. 

\subsection{Strategies for Full-context and Baseline}

Participants reported employing a strategy consisting of \textit{"adding more information"} across both conditions, however, by analyzing the conversation we found that this actually encompassed two distinct strategies. 

\textbf{Strategy 1: Referring to Context} - This strategy prompts the AI Agent to consider the context in which a question is asked. While it showed limited success in the Baseline condition due to the lack of contextual information, its application in the Full Context condition demonstrated a notable impact. For instance, when Participant 2 initially asked about the core capabilities of modern computers, the AI Agent's response was skewed towards quantum computing. However, when the participant refined the question to focus on the context described within the article, the AI Agent adjusted its search within the given context and identified an accurate alternative.

\textbf{Strategy 2: Adding More Information} - This strategy entails providing additional, and specific information to the AI Agent, allowing it to leverage its internal knowledge base. For example,in the Baseline condition, questions about why the chance of decay was 50\% in which the AI Agent had no context led to wrong responses. When "Schrodinger's cat" was mentioned, the AI Agent could tap into its existing knowledge and provide a good response.

\subsection{Design Considerations}
In the realm of AI Agents in multi-window XR environments, our research underscores the critical importance of integrating implicit and explicit inputs to enhance user interactions.To fully enable the integration of explicit and implicit inputs, we propose the alignment between visual attention and verbal input that can enable contextual memory. A new standard for capturing and leveraging user engagement history. This might prove especially relevant in the context of information-rich multi-window environments where there might be multiple contextual targets for an interaction potentially spanning across multiple windows. 


\textbf{Multi-modal input}: developers should focus on harnessing the complementary strengths of visual attention and verbal communication to craft AI agents that are both intuitive and responsive to user input. Visual attention serves as a subtle yet powerful signal for context recognition, allowing the AI to grasp the user's focus without invasive prompts. Concurrently, verbal requests provide a direct avenue for users to convey their needs and queries. 
Our application showcases how directing the AI Agent's attention to user-focused context mitigates the challenges associated with processing extensive inputs. By aligning the AI Agent's focus with the user's, especially in scenarios requiring immediate clarification on recent interactions, the system gains a significant edge over models burdened with the entirety of available context. 

\textbf{Episodic vs Long-term Memory}: our agent contextual memory works by simply discarding older information as the user focuses on new content. The contextual memory in our system has a maximum capacity of 250 words to enable basic episodic memory, rather than long term memory. Changing the capacity  will impact the LLM performances (i.e. responsiveness) and has the risk to run into the same problems we found on the Full Context condition. Therefore, after each request, the contextual memory buffer is cleared, ready to capture new context.

\section{Limitations And Future Work}

A limitation of this system might be related to the underlining eye tracking technology, for example participant behaviour may cause the eye tracker to slip on the participant's head, potentially strongly affecting data quality \cite{Niehorster2020TheTrackers}. Moreover the variability of eye-gaze accuracy across different users, eye traits of ethnic groups may affect the accuracy of eye trackers. For example, a study by  Blignaut and Wium \cite{Blignaut2014} shows that Asian participants' eye-trackers accuracy was worse than that of African and Caucasian participants.  To address these limitations, extended evaluations should consider factors such as individual differences in eye characteristics and the impact of prolonged testing over extended work hours. The current study focused on a scenario which requires the AI Agent's attention aligned with the user's attention to provide clarifications on recently observed content. This approach assumes the user's visual focus aligns with their informational needs. However, scenarios exist where the AI Agent's focus might need to diverge, exploring content the user hasn't directly engaged with—perhaps to highlight overlooked details. Future work should expand beyond this alignment, enabling the AI Agent to consider broader or alternate contexts.

\section{Conclusion}





EmBARDiment, explored a novel framework for integrating context-aware embodied AI agents within extended reality (XR) environments. Our solution goes beyond conventional explicit text-based chat-bots and leverages an \textit{implicit} attention framework to streamline communication and make it more contextually grounded. We utilize the user's gaze and visual saliency to derive relevant context. We ran an experiment comparing three interaction conditions —without context, with full context, and with eye-tracking user-focused context—, participants interacted with an AI agent, adjusting their questions based on the agent's responses.  Findings indicated that using visual attention to guide contextual memory selection resulted in fewer question reformulations and enhanced user satisfaction and perceived helpfulness of the system.This supports our hypothesis that such an approach can enhance interaction efficiency with AI Agents in multi-window XR settings. Our gaze-driven contextual memory extends previous research \cite{Mayer2020EnhancingWorldGaze,romaniak2020nimble,Miniotas2006SpeechGaze}, on multi modal interaction gaze + speech. Additionally our findings highlight the potential to streamline user interactions with AI agents in XR offering specific Design considerations for the design of AI agent functionality in multi-window settings. 


\bibliographystyle{abbrv-doi}

\bibliography{ieeevr.bib}
\end{document}